# Current Practices in the Information Collection for Enterprise Architecture Management

Robert Ehrensperger, Clemens Sauerwein and Ruth Breu
*Institute of Computer Science, University of Innsbruck, Technikerstr. 21A, 6020 Innsbruck, Austria*
*Robert.Ehrensperger@student.uibk.ac.at, {Clemens.Sauerwein, Ruth.Breu}@uibk.ac.at*

Abstract: The digital transformation influences business models, processes, and enterprise IT landscape as a whole. Therefore, business-IT alignment is becoming more important than ever before. Enterprise architecture management (EAM) is designed to support and improve this business-IT alignment. The success of EAM crucially depends on the information available about a company's enterprise architecture, such as infrastructure components, applications, and business processes. This paper discusses the results of a qualitative expert survey with 26 experts in the field of EAM. The goal of this survey was to highlight current practices in the information collection for EAM and identify relevant information from enterprise-external data sources. The results provide a comprehensive overview of collected and utilized information in the industry, including an assessment of the relevance of such information. Furthermore, the results highlight challenges in practice and point out investments that organizations plan in the field of EAM.

## 1 INTRODUCTION

The ongoing digital transformation affects longstanding business models and creates opportunities for new ones (Berman, 2012) that enable increasing the company's profits and sales figures (Amit and Zott, 2012). This digitalization leads to changing business processes and requirements, which force organizations to transform. In the mid-1970s, the average life cycle of very large software applications was between 10 and 15 years, which subsequently decreased to an average value of 5 years by 2005 (Soto-Acosta et al., 2016; Masak, 2006; Beck et al., 2001). This decreasing timespan underlines the fact that organizations continue to transform their enterprise architecture (EA) at an increasingly rapid rate. Business-IT alignment plays a crucial role in this transformation (Roth et al., 2013). It is required to provide transparency between the business requirements and the derived technical implementations. EAM is designed to support and improve this alignment (Maes et al., 2000; Farwick et al., 2016). In particular, it is responsible for transforming a company's "as-is" IT landscape to a "to-be" IT landscape in accordance with an enterprise's business strategy. Thus, the importance of EAM is increasing.

EAM provides a holistic view of the entire enterprise architecture with the help of EA models. These models provide a comprehensive overview of the interrelationships between business processes, applications, processed information objects (e.g., business partner information), and the underlying IT infrastructure components (e.g., server, firewall, network) (Roth et al., 2013).

The success of EAM crucially depends on the amount and quality of available EA information. Thus, various researchers have focused their work on relevant information sources for EAM, such as Farwick et al. (2013) and Buschle et al. (2012).

However, no such studies focus on enterprise-external information sources. Through the introduction of social networks, the internet of things, sensors, and smartphones, the world-wide existing amount of information is significantly increasing (Gantz and Reinsel, 2012). Among them, unstructured information is the fastest-growing type of digital information (Bakshi, 2012). This information provides insights about numerous current changes and events in the real world (Harris and Rea, 2009).

Therefore, this qualitative expert survey investigates the status quo in the information collection and questions which enterprise-external information sources are relevant for EAM and might improve EAM. We designed the following research questions in order to highlight this status quo and to point out the relevance of enterprise-external information for EAM:

- RQ1: What are the current practices of collecting information for EAM in organizations?
- RQ2: What is the most relevant information for EAM?
- RQ3: Which enterprise-external information sources are relevant for EAM in practice?
- RQ4: What is the value of collecting enterprise-external information to EAM?
- RQ5: What are the challenges to collect the identified information?

In order to provide a more detailed understanding of the topic, based on the research questions we derived detailed survey questions according to the recommendation of Gläser and Laudel (2010). In summary, this led to a semi-structured survey containing thirteen questions with twelve sub-questions. In a further step, we selected EAM experts to answer this survey. The results were evaluated according to the method of Mayring (2010).

The remainder of this paper is structured as follows. Section 2 provides a discussion of related work, before section 3 documents the research methodology applied. Section 4 outlines the results, and section 5 discusses the key findings by answering our research questions. Finally, section 6 concludes the research at hand and provides an outlook for future work.

## 2 RELATED WORK

Different authors have focused on leveraging enterprise-internal information sources for EAM. The survey by Farwick et al. (2013) analyzed potential sources and their appropriateness for EAM. Accordingly, they identified many different enterprise-internal information sources such as portfolio management tools, configuration management databases (CMDBs), and license management tools.

Buschle et al. (2012) outlined that an enterprise service bus (ESB) may be used as an appropriate information source for EAM. Furthermore, they showed that leveraging on an ESB leads to an improved quality of EA models.

Only a few researchers have analyzed enterprise-external information sources for EAM. For example, Zimmermann et al. (2017) only highlighted the notion that gathering enterprise-external information may further improve EAM. However, they did not mention any concrete information sources.

The types of enterprise-external information can be diverse. In general, there are three types of information, namely structured, semi-structured, or unstructured (Sint et al., 2009), among which the latter is the fastest-growing type (Bakshi, 2012). Research has shown that more than 80 % of useful business-related information is stored in an unstructured form (Das and Kumar, 2013). This fact underlines the potential that lies in exploiting unstructured information.

The first research steps in gathering unstructured information for specific EAM requirements have been undertaken. For example, Johnson et al. (2016) argued for the use of machine learning techniques to gather unstructured information and maintain EA models. These techniques would even enable handling information with a varying structure. For analyzing massive amounts of diverse EA information (e.g. spreadsheets, documents, presentation), Hacks and Saber (2016) evaluated different big data frameworks. Their goal was to find the best-in-breed solution for the needs of EAM.

Many new opportunities to gather unstructured information such as blog postings, log file contents, or customer reviews are on the rise, although currently research lacks leverage on them. As a first step, it is necessary to identify enterprise-external information sources that may increase the value of EAM. Within the scientific literature, no research could be found investigating the relevance of enterprise-external information sources for EAM.

## 3 RESEARCH METHODOLOGY

This survey is designed in a semi-structured form (Wohlin et al., 2012), containing a mix of open and closed questions. It aims to analyze the current practice of the information collection for EAM in industry.

### 3.1 Participants

In order to identify eligible survey participants, the following selection criteria were defined. Accordingly, we applied the following two criteria for the selection of the survey participants: (1) employment in the field of EAM and (2) at least three years of professional experience in the field of EAM. Moreover, all persons participated voluntarily in this study without any financial compensation.

Table 1: Overview of the participants.

| ID | Role | Years of experience | Industry branch | Region |
|---|---|---|---|---|
| 1 | EAM Lead | 4 | Automotive | GER |
| 2 | Deputy Head of Collaborative EA | 6 | IT Consulting | GER |
| 3 | Management of IT Infrastructure | 18 | Automotive | GER |
| 4 | Enterprise Architect | 10 | Pharma | GER |
| 5 | IT-Architect | 3 | Insurance | - |
| 6 | Senior Expert EA | 10 | Oil & Gas | AT |
| 7 | Head of EAM and Innovation | 12 | Financial Services | GER |
| 8 | Senior Technology Architect | 8 | IT Consulting | GER |
| 9 | Global Technology Consulting | 7 | IT Consulting | GER |
| 10 | Director EAM | 15 | Construction | GER |
| 11 | Head of Digitalization, Strategy, Architecture | 15 | Manufacturing | GER |
| 12 | Project Lead, Product Owner | 9 | Financial Services | GER |
| 13 | Enterprise Architect | 4 | Financial Services | GER |
| 14 | EA Technical Lead | 20 | Defense / Military | GER |
| 15 | Senior Enterprise Business Architect | 5 | Industrial Engineering | GER |
| 16 | Enterprise Architect | 5 | Automotive | GER |
| 17 | Digital Architect | 4 | Financial Services | GER |
| 18 | Senior Enterprise Architect | 20 | Financial Services | GER |
| 19 | Sub Product Owner | 6 | Automotive | GER |
| 20 | IT Senior Professional (IT Referent) | 10 | Automotive | GER |
| 21 | Enterprise Architect, Product Portfolio Manager | 12 | Financial Services / Automotive | GER |
| 22 | Principal Enterprise Architect & Account Chief Architect | 17 | - | GER |
| 23 | Product Portfolio Manager | 4 | Automotive | GER |
| 24 | Enterprise Architect | 10 | Digital Industries | GER |
| 25 | Senior Lead IT Consultant | 14 | Cross-industry | AT |
| 26 | Consultant for Transformation and Business Development | 20 | Information Technology | AT |

Table 1 provides a comprehensive overview of the participants, their roles, years of experience, industry branch, and region. On average, the participants have a work experience of 10.3 years. They work for fifteen different organizations across thirteen different industry branches such as automotive, IT consulting, insurance, pharma, military/defense, financial services, construction, industrial engineering, digital industries, information technology, pharma, oil and gas, and cross-industry. One participant did not disclose the organization's name and another did not state the industry branch. Furthermore, all participants are located in central Europe, although they are working for globally-operating organizations. In order to protect personal dates and keep the organization's intellectual property, the participants are only referenced by an ID. It is worth mentioning that some of the participants work for IT consulting companies. Therefore, they described situations from the perspectives of multiple companies.

### 3.2 Survey Design

The survey contains three different documents. The **preliminary information** contains a description of the research context and the intention of the survey. The **interview protocol** asks personal information like the years of experience in EAM and the industry branch. The **research protocol** lists all survey questions.

In order to highlight the different aspects of the defined research questions, several survey questions were derived from these research questions, according to Gläser and Laudel (2010). Using the survey questions aims to reach a thematic structure of the survey. The idea is to organize the sequence of questions in a way that provides an introduction to the topic and makes it comprehensible for the participants (Kaiser, 2014). This approach facilitates gaining an easier understanding of the survey (Kaiser, 2014). Table 2 shows the research questions and their unique ID's.

Table 2: Mapping of research questions to survey questions.

| RQ ID | Research question |
|---|---|
| RQ1 | What are the current practices of collecting information for EAM in organizations? |
| RQ2 | What is the most relevant information for EAM? |
| RQ3 | Which enterprise-external information sources are relevant for EAM in practice? |
| RQ4 | What is the value of collecting enterprise-external information to EAM? |
| RQ5 | What are the challenges to collect the identified information? |

Moreover, table 3 provides an overview of the survey questions (SQs) and their mapping to the corresponding research questions (RQs).

The survey was conducted between May and November 2019. The surveys were evaluated according to Mayring (2010). This approach was required to analyze the open questions. Mayring (2010) provides a method for qualitative text analysis that offers guidance on how to paraphrase, code terminologies, generalize to a higher abstraction level, and reduce to the core gist of the given answers.

## 4 RESULTS

This section discusses the main results of the survey by answering the research questions.

## 4.1 RQ1: Current Practices of Information Collection for EAM

In order to highlight the current practices of information collection for EAM (**RQ1**), we surveyed the participants about (1) the **information that is currently collected and populated to EA models**, (2) the **automation of the information collection** including an explanation of the current realization approaches and their advantages, and (3) **planned technological improvements in the field of EAM**. Therefore, the survey questions SQ1, SQ3, SQ4, SQ4.1, SQ4.1.1, SQ5, SQ5.1 are answered accordingly.

Our survey discovered the overall distribution of the information that organizations currently collect for EAM. Figure 1 illustrates this distribution, whereby 100 % equals the total number of 26 participants.

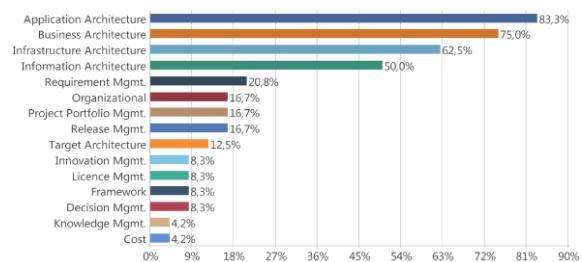

Figure 1: Distribution of the currently-collected EAM information.

Table 3: Overview of survey questions.

| SQ ID | Survey question | RQ ID |
|---|---|---|
| SQ1 | Which information is currently acquired for EAM? | RQ1 |
| SQ2 | Is all relevant acquired information stored within EA models? | RQ2 |
| SQ2.1 | In case the answer to SQ2 is "yes," we asked: What are the most important ones? | RQ2 |
| SQ2.2 | In case the answer to SQ2 is "no," we asked: Which ones are not? | RQ2 |
| SQ3 | How is the EA model information maintenance process designed? | RQ1 |
| SQ4 | Is there already an automatic EA information gathering in place? | RQ1 |
| SQ4.1 | In case the answer to SQ4 is "no," we asked: Would this bring advantages to enterprise architects in your field? | RQ1 |
| SQ4.1.1 | In case the answer to SQ4.1 is "yes," we asked: What are the advantages? | RQ1 |
| SQ5 | Are there any technological improvements planned in the field of EAM? | RQ1 |
| SQ5.1 | In case the answer to SQ5 is "yes," we asked: What are the examples of this? | RQ1 |
| SQ6 | In case that you plan the next project portfolio, what additional information would help you? | RQ3 |
| SQ7 | What additional enterprise-external information may provide added value to EAM? | RQ3 |
| SQ7.1 | What are the advantages for enterprise architects knowing the aforementioned (in SQ7) information? | RQ4 |
| SQ7.2 | What might be the advantages for other stakeholders knowing the aforementioned (in SQ7) information? | RQ4 |
| SQ8 | What is the original source of this information? | RQ3 |
| SQ9 | What are the three main reasons why this information is not already leveraged for EAM? | RQ5 |

The following three examples are the most frequently-mentioned answers in the survey, as shown in figure 1. The majority of the participants (83.3 %) mentioned information objects such as existing interfaces, applications, and their interrelations out of the field application architecture information as currently collected for EAM. The collection of business architecture information such as existing business processes, capabilities, business objects, and business domains was highlighted by 75 % of the participants. Whereby, a business object can be a "customer," and its attributes can be "name," "second name," "age," "country." Furthermore, information that corresponds to the infrastructure architecture such as servers, network devices, deployed technologies was mentioned by 62.5 % of the participants. Half of the participants (50 %) collect information out of the field information architecture, such as business partner information, payment information, or the delivery time information of certain products. 20.8 % of the participants mentioned collecting information for requirement management such as desired changes in certain parts of the EA. It is visible that organizations collect a wide range of information for EAM. 16.7 % of the participants outlined organizational, project portfolio, and release management information. Few participants (12.5 %) stated to collect information about the target EA. A minority of 8.3 % suggested collecting information about innovation management, license management, frameworks, and decision management.

Moreover, information examples were given that are collected only occasionally. Among these, the collection of information about gained EA knowledge (4.2 % of the participants) was stated. In the daily work with EAs, employees gain knowledge such as detailed interdependencies of long-term applications used and their overarching processes. Architects want to access this knowledge gained at the EA models. Moreover, a small minority of participants (4.2 %) expressed the need to collect cost information about (1) business processes, (2) applications, and (3) infrastructure components. Some architects strive to know the total cost of ownership for operating a business process or an application.

In a second step, we focused on the **automation of information collection processes**. Therefore, we asked the participants whether there is already an automatic information collection in place. The vast majority of 73.1 % mentioned that there is no automatic information collection in place. Only 26.9 % stated that they use automatic information collection for EAM.

If there was automatic information collection in place, we continued asking the participants **how this automatic information collection is realized**. The most frequent answers were by using automated data imports out of the (1) CMDB (Configuration Management Database), with automated (2) asset scanning tools, with data imports of (3) middleware, and (4) security tools. The participants highlighted that these imports mainly provide basic master information about technical details such as server names, installation date, or deployed software versions. We asked the participants that do not currently use any automatic information gathering for EAM to assess the **potential of automation**. 77.3 % of the participants who do not have automation in place stated that this would bring advantages to their business. The following **advantages of automatic information collection** were stated:
- Time savings
- Timeliness of information
- Improved collaboration
- Better transparency of the entire IT landscape
- Correctness of information
- Consistency of information
- Accuracy of information

The distribution of the advantages is illustrated in figure 2.

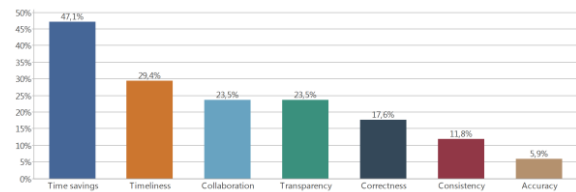

Figure 2: Advantages of an automatic information collection.

When talking about **planned technological improvements within EAM**, more than 62.5 % of the participants stated that they plan technological improvements. Additionally, we asked the remaining participants who **plan technological improvements to list these**. In doing so, investing in the automation of the information collection for EAM was most frequently named by 42.9 % of the remaining participants, whereby some of the participants focused more on the collection of application documents while others more on the collection of business information objects such as business partners, contracts, or bill of materials that are relevant for their business.

## 4.2 RQ2: Relevance of Information

The aim of this study is also to identify the **most relevant information that is acquired and stored within EA models (RQ2)**. Therefore, we discuss the answers related to SQ2, SQ2.1, SQ2.2 within this section.

We asked whether **every relevant acquired information is stored within EA models**, whereby 61.5 % of the surveyed participants responded that not all relevant information is currently acquired and stored within EA models.

Subsequently, we surveyed this remaining 61.5 % of the participants concerning **relevant information that is not acquired and stored within EA models**. The following figure 3 depicts the distribution of the given answers.

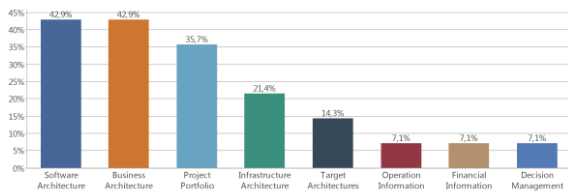

Figure 3: Overview of relevant information that is not acquired and stored within EA models.

Information concerning software architecture such as the application design or information about existing interfaces was named by 42.9 % as missing information. Furthermore, information details out of the business architecture such as business processes, business models and capabilities and the overall business strategy were highlighted by 42.9 % as relevant but not collected. Furthermore, 35.7 % outlined the project portfolio of an organization as often not being collected. The project portfolio contains information about all projects of an organization, with its interrelated dependencies. Having this information linked to different layers of an EA (e.g. conceptual and logical layers) was seen as often missing. Moreover, information about the current infrastructure architecture (21.4 %) and the target EA (14.3 %) was highlighted. Only a few participants (7.1 %) stated operation information, financial information and information about decisions taken as relevant but not acquired and stored within EA models.

In addition, we surveyed the participants who agreed (38.5 %) that **all relevant information is already acquired and stored within EA models** about the **most important information**. The following figure 4 provides an overview of the given answers.

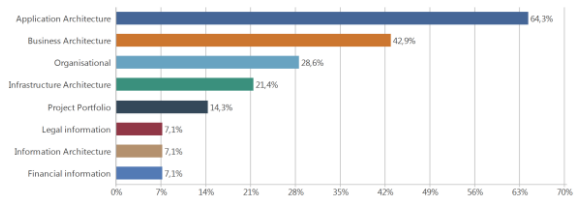

Figure 4: Overview of the most important information.

64.3 % of the remaining participants who agreed that all relevant information is already acquired and stored within EA models outlined examples from the application area of architecture information. For instance, an organization's established interfaces and the need to have an asset inventory that acts as a single point of truth within the EA were expressed as most important. A significant number of participants (42.9 %) also highlighted that business architecture information is most important. As examples of business architecture information, a process inventory, business capabilities, and corresponding IT capabilities were described. A minority of 26.6 % of the participants see organizational information, such as responsibilities, resource allocation, and ongoing activities as most important. Furthermore, infrastructure architecture (21.4 %) and project portfolio information (14.3 %) were expressed as the most important information for some participants. Finally, only a few participants (7.1 %) mentioned examples like legal information, information architecture, and financial information.

## 4.3 RQ3: Relevance of Enterprise-external Information

Besides the relevance of information in general, we also focused on **which enterprise-external information may provide added value to EAM (RQ3)**. We set this focus by addressing the survey questions SQ6, SQ7, SQ8.

The identified enterprise-external information is depicted in figure 5.

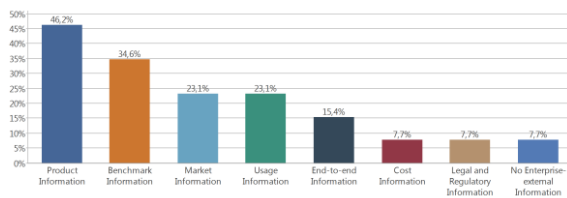

Figure 5: Enterprise-external information that may provide added value to EAM.

The majority of the participants (46.2 %) state product information such as (1) available product versions, (2) product lifecycle information, (3) the functional scope of products, and (4) emerging trends as relevant enterprise-external information. The second most frequent answer (34.6 %) was benchmark information, such as (1) best practice approaches, (2) EA patterns, (3) reference software installations, and (4) touchpoints to external value streams. Few participants (23.1 %) outlined market information, such as (1) buying trends, (2) opinions related to products, (3) financial risk ratings of solution providers.

Moreover, enterprise-external information such as (1) customer feedback about the (2) usage of processes and applications, and information about (3) the unforeseen usage of products is seen as relevant by 23.1 % of the participants. Moreover, 15.4 % of the participants mentioned end-to-end information that involves the entire value chain across company borders such as supplier and customer information about their business processes, business information models, and EA models. A small minority of participants (7.7 %) highlighted cost information, legal and regulatory information, and no enterprise-external information at all as value-adding for EAM.

Furthermore, we investigated the **origin of the identified enterprise-external information.** Therefore, we asked the participants about the origin of their mentioned information examples. Figure 6 illustrates the given answers.

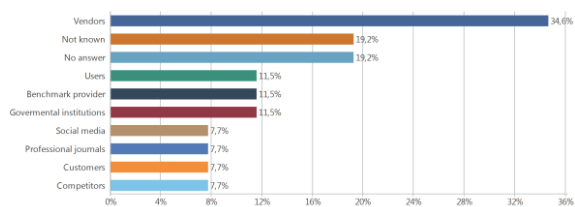

Figure 6: Sources of EAM-relevant enterprise-external information.

The origin is seen mainly with the vendors of certain software products (34.6 %). A significant number of participants responded that either the origin is unknown (19.2 %) or they did not answer this question (19.2 %) within the survey. Further origins were identified at governmental institutions (11.5 %), benchmark providers (11.5 %), and users of the applications (11.5 %). Only a few participants (7.7 %) mentioned social media, professional journals, customers, or competitors as the origin of the given information example.

Moreover, our investigations identified use cases whereby the link between the previously identified origin and the value of information was outlined. For example, Participant (ID 6) stated that information about the product lifecycle underpins enterprise architects to plan the timing for the substitution of products. Further information about the functional scope of available software solution also supports by making fit/gap analysis for this substitution. Both pieces of information have their origin enterprise-externally at vendors.

Furthermore, the participants were also asked **which additional information would help for the project portfolio planning** in organizations.

The functional scope of processes and applications was named by 30.8 % of the participants. Moreover, information about the business models and strategy was stated by 23.1 % of the participants. Additionally, 23.1 % of the participants highlighted data models and flows as relevant additional information for the project portfolio planning. Furthermore, 19.2 % of the participants stated either information about the project organization or did not answer at all. 15.4 % of the participants expressed information about the usage of software solutions and financial information as helpful. Only 7.7 % of the participants mentioned security and compliance information. A small minority of 3.8 % stated product lifecycle information, risk information, information about proof of concepts, information about interrelations between EA layers, and no additional information at all as being helpful for the project portfolio planning.

The stated information examples reveal that for the task of project portfolio planning, information with enterprise-external origin such as the functional scope of applications also plays an essential role besides the enterprise-internal examples.

### 4.4 RQ4: Value of Collecting Enterprise-External Information

In order to assess **the value of collecting enterprise-external information (RQ4)**, we addressed the survey questions SQ7.1, SQ7.2.

Approximately half of the participants (46.2 %) stated that an accurate understanding of future EA needs by being informed about new trends regarding products, technologies, and customer needs would allow enterprise architects to achieve a more active role in designing the EA instead of reacting to demands from top management.

Moreover, the opportunity to conduct comparison-based functional evaluations with other available software products was highlighted as an advantage by 34.6 % of the participants. A comparison-based evaluation may help enterprise architects to identify the best-in-breed software solutions for required EA changes. This value may be leveraged by collecting enterprise-external information about software products (functional scope, supported business processes, best practices) from software vendors.

19 % of the participant reported the benefit of collecting cost information about business processes and the IT systems, whereby the aim is to gain an overview of the return on investment (ROI) of an EA at a glance, allowing enterprise architects to improve their evaluation and decision-making on a target EA. In order to view the ROI, cost information is required that comprises enterprise-external information such as license costs and enterprise-internal information such as the cost of operation.

Furthermore, the survey asked about **advantages for other stakeholders by collecting enterprise-external information**.

The participants most frequently mentioned (38.5 %) an improved decision-making process as an advantage. This advantage is associated with an extended base of information. For example, for business managers this could improve decision-making about investments, the make or buy question, or the prioritization of projects by having information on the end-user perception and utilization of an EA. Moreover, IT managers can more easily re-assess and re-evaluate past decisions about selected standard software products with the help of best practice information concerning applications, business processes, and the re-use of functionalities.

### 4.5 RQ5: Challenges of Collecting the Identified Information

The paper at hand also investigates the **challenges to collect the identified information (RQ5)**. This investigation includes all previously-described examples coming from an enterprise-internal and an enterprise-external environment. We addressed the survey question SQ9 to investigate these challenges. The overall findings are illustrated in the following figure 7.

Approximately one-quarter of the participants (26.9 %) stated that resources play the most challenging role. It was stated that the manual effort to collect the required information is time-consuming that they primarily focus on the collection of the most important information. This survey outlined that organizations perceive the information collection for EAM, not as the most crucial task for their business success. This lack of priority and existing legacy architectures were highlighted as challenging by 23.1 % of the participants. Moreover, we discovered that EA experts suggest missing EA knowledge (19.2 %) and the difficulty of identifying the relevant EAM information (15.4 %) within organizations as a significant challenge. These two points play an essential role in the selection and usage of information.

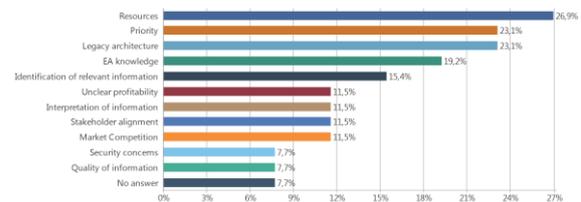

Figure 7: Main challenges why relevant information is not leveraged for EAM.

Furthermore, 11.5 % of the participants stated challenges like unclear profitability of EAM investments, the interpretation of information, the stakeholder alignment, and market competition as challenging. Only a few participants (7.7 %) mentioned examples like concerns about security issues and the quality of acquired information or they did not answer at all.

## 5 DISCUSSION

Within this section, the key findings and potential limitations of our study will be discussed.

### 5.1 Key Findings

**Key Finding 1: Industry does not collect and utilize all relevant enterprise-internal and enterprise-external information for EAM.**

This survey discovered that 61.5 % of the participants stated that not every relevant information is collected for EAM.

We identified a list of not-collected but relevant information for EAM: first, the lack of available information about software architectures (42.9 %) of deployed software solutions as the most important one; second, information about the business strategy, models, and processes is often (42.9 %) not collected and stored within EA models; and third, information about the project portfolio is described by 35.7 % of the participants.

Furthermore, this work discovered an apparent mismatch in assessing the relevance of information. It was revealed that some organizations see information about business processes and capabilities as highly relevant, while others do not collect this information at all (cf. Figure 3). However, this information is essential for achieving alignment between business and IT, which is the primary responsibility of EAM.

This finding emphasizes a gap of relevant information and its collection in practice.

**Key Finding 2: Practitioners highlighted the relevance of enterprise-external information.**

This study discovered the relevance of enterprise-external information for EAM in practice. The participants mentioned several examples of relevant enterprise-external information, relating to available software products and its versions, product lifecycle information, and the functional scope of the products.

For instance, Participant (ID 19) described the relevance of acquiring enterprise-external information by the following use case. Enterprise-external information enables architects to conduct a comparison-based evaluation with other available standard software solutions that focus on the same functional scope. Enterprise architects may focus on the weak parts of the business processes, and try to optimize them by selecting the best-fitting on the market available software solution for this process part.

Besides a list of relevant enterprise-external information, this study also identified the origins of the information. These were seen mainly at vendors of software products (34.6 %), users of applications (11.5 %), benchmarking providers (11.5 %), governmental institutions (11.5 %).

**Key Finding 3: Practice plans to invest in the automation of information collection for EAM.**

The paper at hand identified that industry is planning to direct investments in the automation of information collection for EAM. The automation of the information collection is already a longstanding research topic, as highlighted by many researchers (e.g. Farwick et al., 2011; Moser et al., 2009; Grunow et al., 2013; Buschle et al., 2011). However, the current state in practice is not that far yet. Participants of this study described mostly manual processes to collect and maintain the EA models.

However, the first automated EA information collection processes already exist. Seven participants described automated data imports out of the CMDB, asset scanning, middleware, and security tools. These automated imports collect information from mainly technical layers, such as server names, hardware configurations, and software installations. Thus, this work also identified that no automated collection of business processes and the flow of information objects between IT systems are established within the surveyed industry. Furthermore, our work identified no automated collection of enterprise-external information among the participant's organizations.

Nevertheless, there is a joint agreement on the potential of automation for EAM. 77.3 % of the participants who do not yet have automation in place agreed on the potential of automation of the information collection processes. Furthermore, this paper has outlined the advantages observed, such as time savings, better timeliness of information, and improved collaboration.

Moreover, the majority of the described organizations (62.5 %) plan technological improvements in the field of EAM. Many of them (30 %) focus on the automation of the information collection processes. Participants highlighted this investment as the most pressing one in the field of EAM. This finding also reveals that the potential of automation is not yet fully leveraged.

**Key Finding 4: Current practice has only limited resources for the information collecting of EAM.**

Our work identified a list of challenges why EAM-relevant information is not collected. The first three challenges are a lack of resources, missing priority, and EA knowledge. These challenges are closely linked to each other and may have the same origin.

Organizations may tackle these three challenges by assigning more budget to EAM projects and EA employees. Accordingly, it is essential to focus on the return of their investments. In the case of collecting information for EAM, it is difficult to provide a method that calculates reliable the expected ROI. However, organizations need to be able to calculate upfront a clear business case for investing in EAM.
As a result of this, research can provide guidance on the assessment of business cases by providing statistical information on the return of investments from other EAM projects.

## 5.2 Limitations

Our survey might be limited by certain threats to validity, namely the (i) selection of eligible participants, (ii) the missing reproducibility of the results, and (iii) false categorization and analysis.

In order to overcome (i), we applied the following two participant selection criteria: (1) employment in the field of EAM and (2) at least three years of professional experience in the field of EAM. A detailed description of the participants can be found in section 3.1.

In order to overcome (ii), we noted the personal information, including contact details of each participant and we, used the software tool MAXQDA (Rädiker and Kuckartz, 2019) for the data analysis. This tool provides traceability from given survey answers to the analysis results and the conclusions that we draw.

We evaluated the survey according to a method for qualitative text analysis introduced by Mayring (2010). This method provides systematic guidance on how to paraphrase, code terminologies, generalize to a higher abstraction level and reduce to the core gist. Moreover, each instance of the paraphrasing and coding was reviewed by at least two authors of this publication. As a result, the risk of (iii) is at an acceptable level.

# 6 CONCLUSION AND OUTLOOK

EAM's principal objective is to optimize the strategic IT alignment of organizations. A thriving EAM crucially depends on available information within the EA models. Therefore, the information selection and collection is a pivotal issue.

In this paper, we analyzed the current practices of the information collection for EAM in the industry within Europe. Initially, we looked at the related work and discovered that (1) the automation of information collection for EAM is already a longstanding discussed topic within research, although current practices are not investigated at all, and (2) only little research has taken place in the field of collecting enterprise-external information for EAM. Subsequently, we conducted a qualitative expert survey among EAM practitioners to address the research gaps (1) and (2).

Our survey reveals that the industry within Europe does not collect all relevant information, while EA practitioners underline the utility value of this information for their organizations. Furthermore, we discovered that EA practitioners also express the relevance of enterprise-external information for EAM. Moreover, we could outline an emerging trend since most organizations lack but plan to invest in the automation of information collection for EAM. Finally, we also identified the main challenges of leveraging all relevant information for EAM. Our results provide researchers with a detailed view of the current practices in information collection for EAM.

The findings of this survey rise to several directions for further research. The lack of automation of the collection of information, such as business processes, business information objects. Future research could highlight how to **automate a semantical integration into EA models** of these information examples. In terms of the challenges identified, further research could give guidance on the **assessment of investments within EAM** concerning the ROI. Finally, regarding the collection of enterprise-external information, further research may investigate **frameworks that enable integrating external sources** into an EA model.